\documentclass[a4paper]{mem}
\usepackage{natbib}
\usepackage{graphicx}
\begin{document}
   \title{Relativistic Outflows \\
     from Active Galactic Nuclei
%\thanks{supported by DFG Sonderforschungsbereich 439}
}

   \author{M. Camenzind \inst{1}\fnmsep
%\thanks{this is a place for placing a footnote in the author field }
}

\offprints{M. Camenzind}
%\mail{Landessternwarte K\"onigstuhl, D-69117 Heidelberg }

   \institute{Landessternwarte K\"onigstuhl,
K\"onigstuhl 12, D-69117 Heidelberg \\
\email{M.Camenzind@lsw.uni-heidelberg.de}\\
             }

   \abstract{
We review recent progress in the theory of relativistic jet production in AGN and Micro-quasars. The current popular model for launching, accelerating and collimating astrophysical jets is based on magnetohydrodynamics (MHD). AGN jets are most probably powered by energy extracted from either an accretion disk or a rotating Black Hole. A strong electromagnetic field in the central engine, coupled with differential rotation, serves to convert rotational energy into kinetic energy of outflows. In the last few years, some progress in understanding accretion processes for rotating Black Holes has been made: standard disks are truncated at some radius depending on the accretion rate so that disk outflows are driven away by the hot inner disk. Slow outflows are the norm when the magnetorotational instability is at work in the weak field limit. In order to achieve Lorentz bulk factors of about 10, strong large-scale magnetic fields must thread the Black Hole's ergosphere. The production of relativistic ouflows is then completely understood within stationary MHD models. Some special cases, such as the force-free limit (so-called Poynting flux jets) have been investigated in the last years. In general, however, the jet plasma not only consists of magnetic fields, but also of thermal ions and electrons. In the collimated region, the electrons are boosted up to a non-thermal distribution, which is the basis of all observations.
   }
   \authorrunning{M. Camenzind}
   \titlerunning{Relativistic Outflows}
   \maketitle
%
%________________________________________________________________

\section{Introduction}

A Black Hole is an unavoidable component of any massive galaxy, and as such the ultimate driver for accretion towards the center, provided there is sufficient fuel available on the parsec--scale. The mass of these Black Holes scales with the mass of the spheroidal component of the host galaxy, or with the depth of its gravitational potential (so-called Magorrian relation) -- the most massive Black Holes are harbored by the most massive galaxies, i.e. the elliptical ones. But only a minority of accreting Black Holes is able to launch relativistic outflows. These supermassive Black Holes have probably been formed by stellar collapse at redshifts $\simeq 15$, and then they have grown in mass and angular momentum by merging and accretion processes on cosmological time--scales.

Similar to rapidly rotating neutron stars, a Black Hole with a non--vanishing Kerr parameter disposes of some rotational energy, and this can be extracted by means of macroscopic electrodynamic processes (\citet{Cam03}).
Since the invention of the Blandford--Znajek process, the formation of the MHD jet, in a disk threaded by large-scale poloidal fields, has been studied extensively by many groups.

In the following, I review recent progress in the theory of relativistic jet production. The presently favored mechanism is an electrodynamic one, in which hot plasma is accelerated by Lorentz forces that are generated by rotating magnetic fields. The most pressing issues of current interest are understanding what factors control the jet power, its speed, and its degree of collimation, and how these properties determine the type of jet observed and its effect on its environment. Recent observations of microquasars, pulsars, gamma-ray bursts (GRBs) and core-collapse supernovae indicate that jets play an important role everywhere.

%
%                                                Two column figure
%----------------------------------------------------------- S_vib
%   \begin{figure}
%   \centering
%   \resizebox{\hsize}{!}{\includegraphics{figura3_articolo.eps}
%   \includegraphics{figura4_articolo.eps}}
%   \caption{Left panel: Best fit of the cluster with
%            1.75 Gyr theoretical isochrone with OPAL EOS.
%            Right panel: Best fit of the cluster with 1.95 Gyr
%            theoretical isochrone with Straniero (1998) EOS}
%     \label{fit3}
%\end{figure}
%
%______________________________________________________________

%                                     Two column figure (place early!)
%______________________________________________ Gamma_1 (lg rho, lg e)
   \begin{figure*}
   \centering
%   \resizebox{\hsize}{!}{\rotatebox[]{0}{\includegraphics{CamenzindF1.ps}}}
   \includegraphics[angle=0,width=12.0cm]{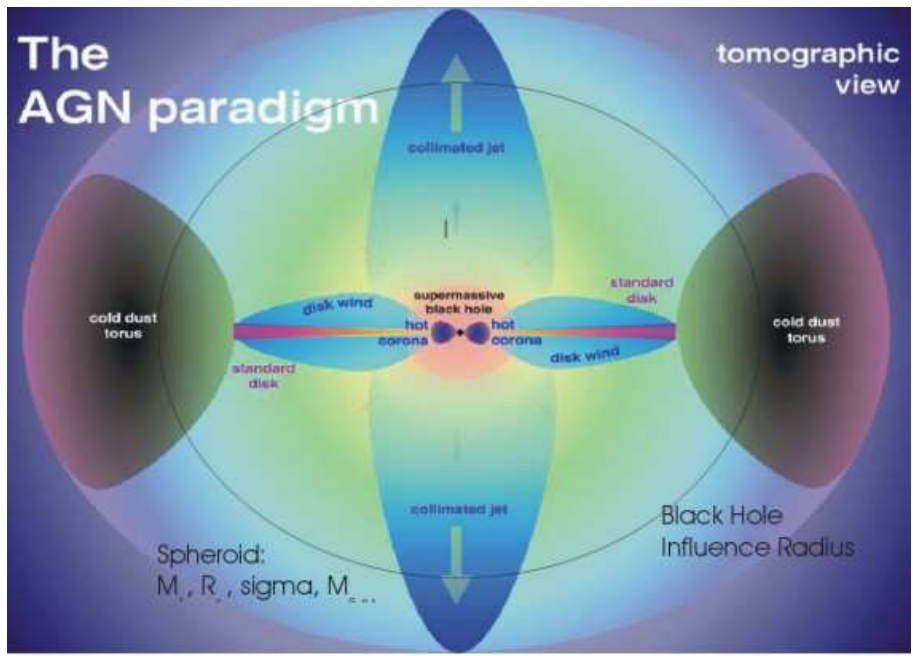}
   \caption{Tomographic view of the parsec--scale structure of active galactic         nuclei (\citet{Mueller04}). Depending on the age of the nuclear spheroid,
   mass injected by stars in the spheroid fill in a kind of torus structure which      contains the fuel for the accretion process (as in 3C 273). In elliptical
   galaxies of the local
   Universe, not much gas is left over, so that a hot ISM fills up the central
   spheroid (as in M 87 e.g.).
   }
              \label{agnparadigm}%
    \end{figure*}
%______________________________________________________________
%

\section {Outflows and the AGN--Paradigm}
Active galactic nuclei (AGN) may be defined loosely as the central regions of galaxies which show substantial energy release beyond what can be attributed to normal processes from stars, ISM, and their interactions. This includes Seyfert galaxies, radio galaxies, quasars, QSOs, BL Lac objects, and possibly the common LINERs in ordinary galaxies.
Black holes with masses of a million to a few billion times the mass of the Sun are believed to be the engines that power nuclear activity in galaxies (Fig. \ref{agnparadigm}). Active nuclei range from faint, compact radio sources like that in M31 to quasars like 3C 273 that are brighter than the whole galaxy in which they live.

The most plausible mechanisms for energizing large, powerful extragalactic sources from AGNs involve well-collimated outflows near the symmetry axes of accretion disks around black holes. The Black Hole's potential well is needed (i) to generate the observed luminosity at high efficiency and (ii) to provide the relativistic outflow velocities implied by super-luminal proper motions on parsec scales.

Outflows generated from accretion disks are highly collimated already on the parsec--scale (see M 87 or Cen A). As elliptical galaxies in the local Universe are filled up with a hot interstellar gas with typical densities $n_e < 1$ cm$^{-3}$, the outflow is highly ballistic on these scales with a density contrast $\eta_{pc} = \rho_b/\rho_{ISM} \simeq 100$. However, due to the rapid expansion of the jets, $\rho_b \propto 1/R^2$, and a roughly constant density in the intracluster gas, the density contrast on the kiloparsec--scale is then much smaller than unity ($\eta_\mathrm{kpc} \simeq 10^{-3}$ in Cyg A on the scale of one kpc).

\section{Truncation and Jet Launching}
\begin{figure*}
   \centering
   \resizebox{\hsize}{!}{\rotatebox[]{0}{\includegraphics{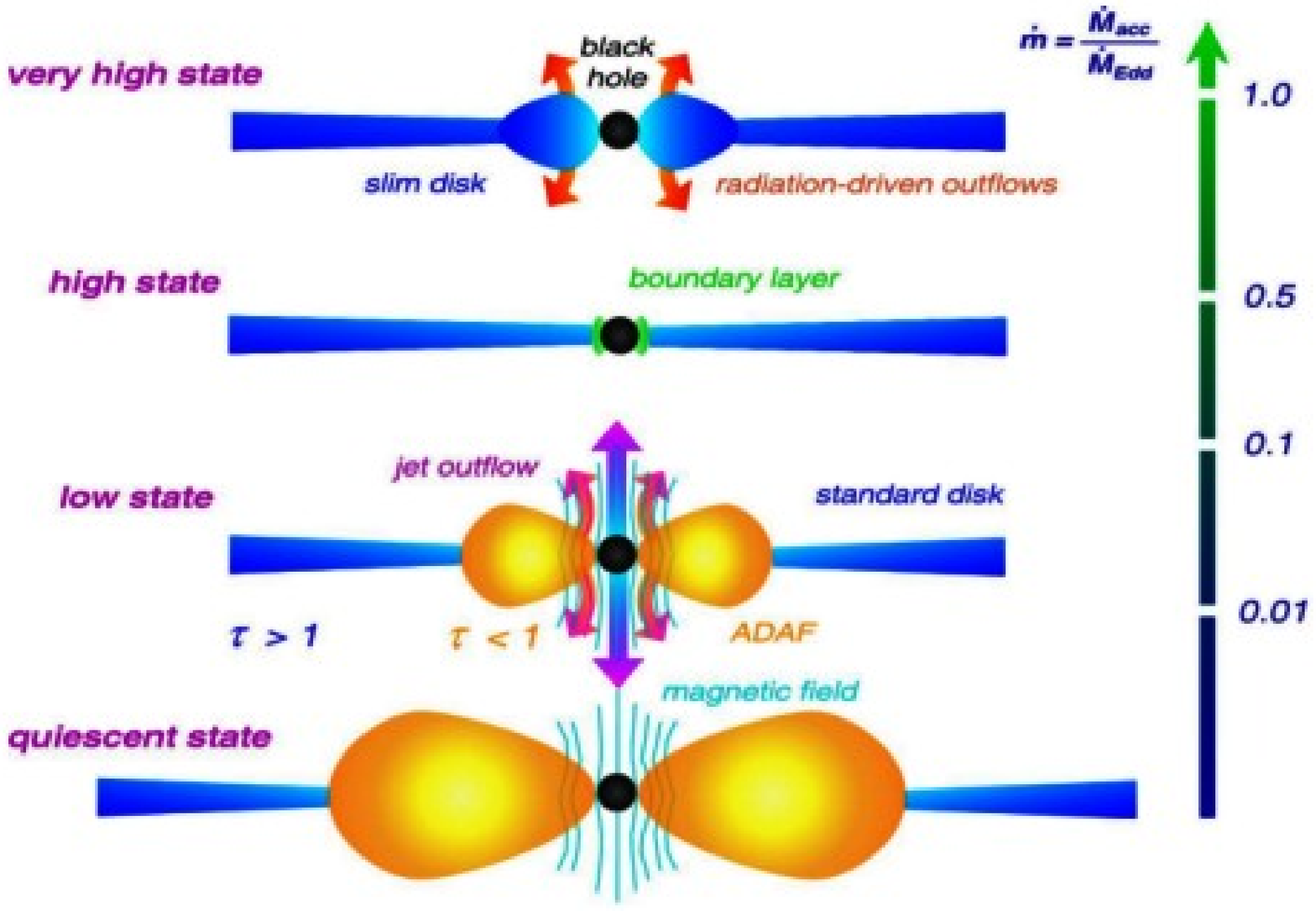}}}
   \caption{Illustration of the formation of jets in accretion theory (\citet{Mueller04}). Under the high accretion rates found in the early Universe, a standard accretion disk (SDD) is formed far away from the central Black Hole, which is truncated at a certain radius near the marginal stable orbit of the Black Hole. Here, a transition occurs towards an inner advection dominated disk which is of toroidal shape. The standard disk is the origin for the UV--bump observed in quasars, while in the hot inner disk hard X--rays are produced by Compton scattering of soft photons.  }
    \label{truncation}%
    \end{figure*}

X--ray binaries thought to contain Black Holes show at least {three spectral states} (\cite{Fender04}, Fig. \ref{truncation}). The {\bf low/hard state} (LHS) is characterized by a power--law and an exponential cutoff at about 200 keV or a thermal Comptonisation model with an electron temperature of about 70 keV. A {quiescent state} has been suggested to exist, but in reality, the X--ray and radio properties of low luminosity BH candidates
%                                     Two column figure (place early!)
%______________________________________________ Gamma_1 (lg rho, lg e)
%______________________________________________________________
%
form a continuum down to the lowest observable luminosities. The spectrum of the {\bf high/soft state}  (HS) is dominated by a thermal component thought to arise in a standard thin disk (SSD). This state also exhibits a weak power--law tail without observable cutoff. The very {\bf high state} (VHS) displays the same two components, but with the steep power--law dominating the total flux. Steady radio emission, which is likely coming from a jet, has been found in the low/hard state, and strong radio flares have been seen during the very high state. No strong radio emission has ever been detected in the high/soft state.

\subsection{Truncated Accretion}
The most broadly accepted view on the strucuture of these disks is that the character of the flow is mostly determined by the Eddington ratio $\dot m = \dot M/\dot M_{Edd}$. In high Eddignton ratio objects, accretion proceeds through a cold optically thick disk, while for lower Eddington ratio the cold disk evaporates close to the BH. Below a cetrain radius $r_{tr}$, depending on the accretion rate $\dot m$, the accretion flow proceeds through some form of optically thin hot flow. A plausible geometry is shown in Fig. \ref{truncation}.

For accretion rates above a few percent Eddington, accretion disks on scale much beyond the marginal stable orbit are well modeled by standard disk models (Fig. \ref{truncation}). These disks vertically collapse due to efficient radiative cooling which is the origin of the observed UV--bumps in quasars. These disks will however be truncated at some radius $r_{tr}$, whose position is dictated by the overall accretion rate and the nature of the turbulence which is responsible for the angular momentum transport (\citet{Gracia03}). Here the disk is transformed into a hot geometrically thick and optically thin disk (sometimes called ADAF, or hot accretion torus). In this part, energy is mostly advected inwards and disappears inside the horizon of the black hole. In this part, the hard X--rays are generated by Compton scattering of soft photons.

\paragraph{Fundamental Time--Scales:}
If the accretion flow is roughly Keplerian, the most fundamental time--scale is the dynamical one given by the Keplerina frequency,
%$t_\mathrm{dyn} = 2\pi/\Omega_K(R) = \sqrt{ R^3/(GM_H) }$.
which can conveniently be expressed in suitable units
\begin{equation}
  t_\mathrm{dyn} = 1.05\,\mathrm{days}\,M_{H,9}\,(R/R_S)^{3/2} \,.
\end{equation}
Thermal time--scales are somewhat larger due to the turbulence parameter $\alpha$,
$t_\mathrm{therm} = t_\mathrm{dyn}/\alpha$.
When the transition radius changes with time, we need an estimate for the characteristic time--scale of the removal of the cold disk from a given radius. This removal occurs in a change of the accretion flow into an optically thin flow
$t_\mathrm{evap} = E/(\eta\dot Mc^2)$
with $E = \pi R^2\Sigma\,kT/m_p$, which can be expressed as
\begin{equation}
  t_\mathrm{evap} \simeq 10\,\mathrm{yrs}\,\dot m_{0.1}\,M_{H,9}\,
    (R/10\,R_S)^2 \,.
\end{equation}

In radio--loud quasars, knots are ejected with a cyclic time--scale of a few years (in 3C 273 one per year, in 3C 345 every fifth year). It is tempting to relate this time--scale to the above evaporation time--scale. Since the mass in 3C 273 is probably similar to the Black Hole mass found in M 87, where $M_H = 3\times 10^9\,M_\odot$, one year is close to the evaporation time for a radius of a few Schwarzschild radii. Since 3C 273 is a bright quasar, truncation must occur very near to the marginal stable orbit. The mass derived for 3C 120 from reverberation is $M_H = 3\times 10^7\,M_\odot$ and this source is probably in the low state.

\citet{Hujeirat04} presents a self--similar solution for the 3D axi-symmetric radiative MHD equations, which revisits the formation and acceleration of accretion-powered jets in AGNs and microquasars.

\subsection{Non--Radiative Accretion Flows and Jet Launching}
Recently, magnetic fields play an important role for driving the turbulence in the accretion disk since the Maxwell stress coupled with magnetorotational instability (MRI: \citet{Balbus91, Balbus98, Balbus04}) is the only source of the efficient angular momentum transport. Therefore, magnetic fields in the disk may be a plausible source of the large--scale magnetic fields and three dimensional (3D) MHD simulations are inevitable.
These inner advection--dominated flows around Black Holes can now be simulated including relativistic effects (\citet{Krolik}). Turbulence, which is the source of the angular momentum transport, is excited by the magnetorotational instability (MRI) in weakly magnetized plasmas. Typically, the gas pressure initially dominates the mangetic pressure by at least a factor of one hundred. A torus is a suitable initial condition, though this does not allow to study the quasi--stationary state of the turbulence.
%
%_____________________________________________________________
%                                    One column rotated figure
%-------------------------------------------------------------
   \begin{figure}
   \centering
   \includegraphics[angle=0,width=6.5cm]{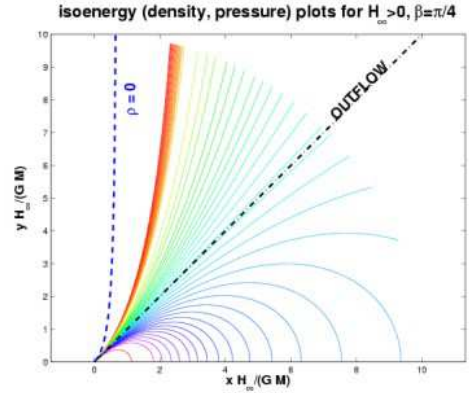}
      \caption{Contours of equipotential surfaces for the case, when $\Omega$
      is 0.84 of its Keplerian value (\citet{Balbus04}). The separatrix marked              OUTFLOW separates inflow from outflow regions. Open contours become
      very closely packe dnear the $\rho = 0$ boundary.
              }
         \label{Torus}
   \end{figure}
%
%_____________________________________________________________

In principle, in a torus the rotation profile is a free function, but all MRI simulations show that a Keplerian distribution is rapidly established. This is due to a rigorous outward angular momentum transport resulting from the MRI. Near a Black Hole, a kind of adiabatic geometrically thick hot disk (temperature $\simeq 10^{12}$ K) is always established (Fig. \ref{Torus}). In the inner part, plasma from the torus is driven away along the open contour surfaces in a cone with half opening angle of roughly 40 degrees (\citet{DeVilliers03}). Indeed simulations show a pile--up of plasma close to the last open equipotential surface. These are thermal outflows which will not be accelerated to relativistic speeds.

\section{The Launching of Relativistic Jets}
Therefore, the question remains under which conditions relativistic collimated outflows can occur. In the following we discuss two scenarios which show collimation by magnetic effects. The first example has been worked out only in the non--relativistic regime.

\subsection{Jets Formed in Accretion--Ejection Structures}
In contrast to the above studies, \citet{Casse02, Casse04} recently presented MHD computations demonstrating the launch of superfast magnetosonic collimated jets from a resistive accretion disk. This is achieved by fully accounting by an appropriate energy equation in the MHD model of the accretion disk with large--scale magnetic fields
\begin{eqnarray}
  {{\partial e}\over{\partial t}} &+&
    \nabla\cdot\Biggl[ \vec{v}\left(e + P + {B^2\over{8\pi}}\right) -
      (\vec{v}\cdot\vec{B})\vec{B} \Biggr] \nonumber \\
     &=& \eta\,\vec{J}^2 -
      \vec{B}\cdot(\nabla\times \eta\vec{J}) \,.
\end{eqnarray}
Resistivity $\eta$ is only operating in the disk and provides a local Joule heating.
The typical temperature obtained near a BH is then given as
\begin{equation}
  T_0 = {{\mu m_p\,\Omega_K^2H^2}\over k_B} =
     10^{11}\,K\,\left( {R\over R_S} \right)^{-1} \,.
\end{equation}
%This temperature scaling leads to the density scaling for M 87 type galaxies
%\begin{eqnarray}
%  \rho_0 &=& {{\dot M_a}\over{4\pi V_RRH}} =
%    2.7\times 10^{-13}\,\mathrm{g\,cm}^{-3} \nonumber \\
%      &&\left( {{\dot M_a}\over{M_\odot\,yr^{-1}}}\right)
%      \left( {M_H\over{10^9\,M_\odot}} \right)^{-2}
%      \left( {R\over R_S} \right)^{-3/2} \,.
%\end{eqnarray}
%The corresponding pressure is then given as
%\begin{eqnarray}
%  P_0 &=& \rho_0\Omega_K^2H^2 =
%    1.2\times 10^6\,\mathrm{cgs}\nonumber \\
%      &&\left( {{\dot M_a}\over{M_\odot\,yr^{-1}}}\right)
%      \left( {M_H\over{10^9\,M_\odot}} \right)^{-2}
%      \left( {R\over R_S} \right)^{-5/2} \,.
%\end{eqnarray}
In standard disks, density and pressure scale directly with the accretion rate $\dot M_a$. Typical magnetic field strengths are then given by equipartition arguments (higher field strengths would lead to magnetic pinching of the disk)
\begin{eqnarray}
  B_0 &=& 3.9\,\mathrm{kG}\,\sqrt{\beta_p}\,
      \left( {{\dot M_a}\over{M_\odot\,yr^{-1}}}\right)^{1/2}\nonumber \\
      &&\left( {M_H\over{10^9\,M_\odot}} \right)^{-1}
      \left( {R\over R_S} \right)^{-5/4} \,.
\end{eqnarray}
For Black Holes, magnetic fields always scale with one over the square root of the mass, whenever the accretion rate is expressed in relative units $\dot m$. Stellar Black Holes have therefore field strengths that are of the order of 10 Mega--Gauss under high accretion rates.
%
%_____________________________________________________________
%                                    One column rotated figure
%-------------------------------------------------------------
   \begin{figure}
   \centering
   \includegraphics[angle=0,width=6.5cm]{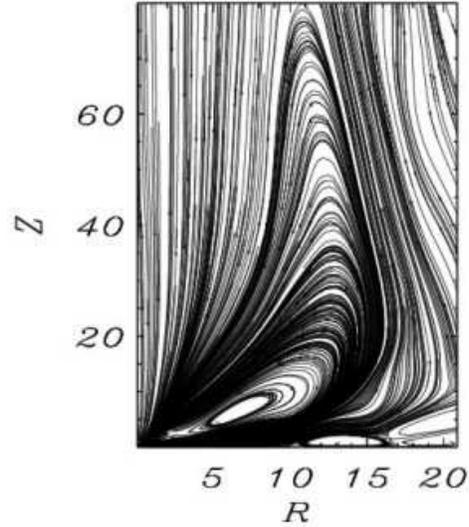}
      \caption{Current distribution in the meridional plane of jets launched by
      a quasi--Keplerian disk (\citet{Casse04}). Closed current loops are
      responsible for the collimation of the disk outflows at high vertical
      distance above the disk surface.
              }
         \label{Currents}
   \end{figure}
%
%_____________________________________________________________

The result of numerically simulating the evolution of the toroidal magnetic structure under resistive accretion is displayed in Fig. \ref{Currents} (\citet{Casse04}). The initial configuration is close to a radial equilibrium where Keplerian rotation balances gravity of the central object. The rotation of disk matter initially twists the purely poloidal field lines such that outflow is driven away from the disk surface. This outflow is less dense than the disk, but much denser than the original corona. The configuration attains a new equilibrium, as time increases, where the poloidal velocity becomes parallel to the poloidal magnetic field, as expected in stationary models. In addition, the outflow is {\bf radially collimated}, and crosses the Alfven und fast magnetosonic points to become super--fast--magnetosonic, before it crosses the outer boundary.

\subsection{Jets and Ergospheric Flux Tubes}
The MRI examples now tell us probably how plasma can be injected into a conical structure, and the above example demonstrates the ability of magnetized outflows to collimate into a nearly cylindrical structure. The most efficient shearing of magnetic fields occurs in the ergosphere of a rapidly rotating Kerr Black Hole, where plasma rotation is dictated by the frame--dragging effect (represented by the frame--dragging potential $\omega(r,\theta)$)
\begin{equation}
  \Omega = \omega + {\alpha^2\over R^2}{\lambda\over{1 - \omega\lambda}} \,.
\end{equation}
For any given angular momentum distribution $\lambda$ in the hot disk, the redshift factor $\alpha$ of the Kerr metric, which vanishes at the horizon, reduces its influence on the plasma rotation so that within the ergosphere only frame--dragging is relevant. In distinction to the above example, currents $T\propto RB_\phi$ are driven by the differential rotation of absolute space as is shown by the evolution equation for the toroidal magnetic field in the infinite conductivity limit (\citet{Cam98, Cam03})
\begin{eqnarray}
  {{\partial T}\over{\alpha\,\partial t}} &+& (\vec{v}_p\cdot\nabla) T
    - R^2\nabla\cdot(T\vec{v}_p/R^2) \nonumber \\
    &=& R^2(\vec{B}_p\cdot\nabla)\Omega \,.
\end{eqnarray}
Except for the correction by the redshift factor $\alpha$, this equation is exactly identical to its Newtonian counterpart. Therefore, currents are driven, where shear, $\nabla\Omega$, is maximal, and this is in the ergosphere. This process of winding up magnetic flux tubes in the ergosphere has recently been demonstrated by \citet{Semenov}. Currents are not closing to the horizon of the Black Hole, as required by the original Blandford--Znajek process, but they close within the ergosphere. It is the differential rotation of absolute space which drives the enormous current loops of relativistic jets. This process produces a Poynting flux with a magnetic luminosity given by
$L_\mathrm{mag} = \Omega_*I_*\Psi_\mathrm{ergo}/c$,
where $\Omega_* < \Omega(R_*)$ is the rotation of the field line at footpoint $R_*$ and $\Psi_\mathrm{ergo}$ is the magnetic flux covered by the ergosphere. An upper limit to the field rotation is given by the rotation of the horizon $\Omega_H = \omega(r_H)$
\begin{equation}
  \Omega_* < \Omega_H \simeq
  10^{-4}\,\mathrm{rad\,s}^{-1}\,(10^9\,M_\odot/M_H) \,.
\end{equation}
Maximum luminosity is then found for the parameters (\citet{Cam98})
\begin{eqnarray}
  L_\mathrm{mag} &\simeq& 3\times 10^{46}\,\mathrm{erg\,s}^{-1}
    {{10^9\,M_\odot}\over M_H} \nonumber \\
    &\times& {\Psi_\mathrm{ergo}\over{10^{33}\,\mathrm{G\,cm}^2}}\,
    {I_*\over{10^{18}\,\mathrm{Amps}}}\,.
\end{eqnarray}
This is sufficient to explain the highest kinetic jet luminosities.

Plasma injected into these ergospheric flux tubes is then accelerated when it flows along converging flux tubes by transforming Poynting flux into kinetic energy.
Relativistic outflows result when the magnetisation of the hot coronal plasma is sufficiently high, expressed in terms of the Michel parameter $\sigma$ (\citet{Cam98})
\begin{equation}
  \sigma_* = {{B_{p,*}R_*^2c}\over{4\pi\mu\eta R_L^2}} \simeq
     {{B_{p,*}^2R_*^4}\over{4\dot M_\mathrm{Jet}c R_L^2}} \,,
\end{equation}
where $R_L = c/\Omega_*$ denotes the light cylinder of the flux tube. For rapidly rotating Black Holes, the light cylinder is bounded by $R_L > c/\Omega_H = 2r_H/a_H > 2r_H$. A minimal light cylinder for extreme Kerr is then about two Schwarzschild radii.

It is important to realize that, if an MHD mechanism for jet
acceleration is adopted, then this implies that (at least initially)
the jets so-produced must be Poynting flux--dominated (PFD).  By definition,
$\Gamma_\infty >>1$ implies that the kinetic energy greatly exceeds the rest
mass--energy of the flow.  And, for an MHD jet, the final velocity is
at least of order the Alfv\'en speed, so the Alfv\'en Lorentz factor also
must be large
\begin{equation}
  \Gamma_{A}^2 \equiv {V_{A}^2\over c^2} = {B_*^2\over{4 \pi \rho_* c^2}}
   \simeq {{B_*^2 R_\mathrm{ergo}^2}\over{4\pi\dot M_Jc}} > 1 \,.
\end{equation}
That is, the flux tubes must have low mass--loading, and the energy
flow must be dominated by the flow of electromagnetic energy (Poynting
flux), not kinetic energy. As the flow accelerates, Poynting flux
is slowly converted into kinetic energy flux, until the two are of the
same order of magnitude (\cite{Cam98,VK01,V04}).  Eventually, mass entrainment
from the interstellar medium can increase the baryon loading.

\section{On Collimation and Jet Plasma}
No code available on the market is yet able to follow the evolution of plasma outflows from the immediate vicinity of the horizon into the domain where collimation occurs, i.e. to scales of hundreds to thousands of gravitational radii. In these domains, we are still urged to use the methods of stationary special relativistic MHD \citet{Cam98}. These methods have the advantage that we can understand the acceleration of plasma to relativistic speeds and also the question of collimation.

%\paragraph{Collimation:}
As pointed out by many authors, there are both theoretical
and observational reasons for concluding that slow acceleration
and {\bf collimation} is probably the norm for jet outflows in these
sources.  Non-relativistic and relativistic
models of MHD wind outflows attain solutions where the wind
opening angle is wide near the accretion disk and then narrows
slowly over several orders of magnitude in distance from the disk.
Furthermore, recent observations of M87, for example, by \cite{JBL99}
suggest that the opening angle of the jet is more than $60\deg$ at
the base, collimating to a few degrees only after a few hundred
Schwarzschild radii.  Furthermore, the lack of significant 'Sikora'
bump in the X-ray light of most radio quasars indicates that the
flow at the base of most quasar jets must also be broad and probably
sub-relativistic, only accelerating to relativistic flow much further
from the black hole.

%\paragraph{Nature of the Plasma in Relativistic Jets:}
According to the previously discussed phenomena, the plasma in jets is normal {\bf ion--electron plasma}, i.e. just disk plasma. Since MHD jets are prone to internal shocks (pinch and kink modes), the electrons are always heated to relativistic energies, while the protons probably stay sub--relativistic. Due to various acceleration mechanisms, the electron energy distribution attains a bumpy structure with a kind of quasi--thermal bump given by a minimal Lorentz factor $\simeq 50 - 100$ and a high energy tail. This energy spectrum is responsible for the gamma--ray emission observed in radio--loud quasars (3C 273 e.g.), where IR photons from hot dust are scattered by the relativistic component.

%\section{Propagation on the Large Scales}

%
%                                                Two column figure
%----------------------------------------------------------- S_vib
%   \begin{figure}
%   \centering
%   \resizebox{\hsize}{!}{\includegraphics{figura5a_articolo.eps}
%   \includegraphics{figura5b_articolo.eps}}
%   \caption{Left panel: solid line: theoretical standard (without overshoot)
%            2.0 Gyr isochrone with Straniero (1998 \cite{straniero}) EOS (Z=0.007 Y=0.244) with
%            overimposed the corresponding synthetic CM diagram including 30\% of
%            binaries; dashed line: theoretical 2.2 Gyr isochrone with Straniero
%           (1998\cite{straniero}) EOS and mild overshooting (${l}_{ov}=0.1
%           {H}_{p}$).
%          Right panel: NGC2420 CM diagram by Anthony-Twarog et al. 1990 \cite{anthony-twarog}.
%           Observational data are not dereddened.
%     \label{fit1}
%\end{figure}
%
%______________________________________________________________

\section{Conclusions}
In the review above I have emphasized several important points
in the study of relativistic jets:
High energy jet sources of all types should be considered
when attempting to understand relativistic jets.  Micro- and
Macroquasars both provide important clues to the mechanisms at
work.
There are observational reasons for believing that the same
source may produce jets of rather different Lorentz factors, either
simultaneously or in different accretion states.
Similarly, there are natural theoretical reasons for believing
that more than one MHD jet launching mechanism may be at work in
a given Black Hole engine.
Some jet production mechanisms at work near the black hole
rely on the extraction of black hole rotational energy.  This
provides a third parameter, in addition to Black Hole mass and
accretion rate, that potentially can explain why sources with
similar optical appearance are radio loud and some are radio quiet.
It is no longer reasonable to consider accretion
models without also considering jet production as an integral part
of the accretion process.  Many, if not all, sources produce jets, and
it is clear that the production of a jet is affected by, and can affect,
the structure of the accretion flow.

\begin{acknowledgements}
      Part of this work was supported by the German
      \emph{Deut\-sche For\-schungs\-ge\-mein\-schaft, DFG\/}
      Sonderforschungsbereich 439
\end{acknowledgements}

\bibliographystyle{aa}

\end{document}